\documentclass[manuscript]{aastex62}
\usepackage{amsmath}
\usepackage{graphics}
\usepackage{natbib}
\usepackage{amsmath}
\usepackage{amssymb}
\usepackage{longtable,booktabs}
\usepackage{array, makecell}

\begin{document}
\title{Identification of Magnetic Flux Ropes from Parker Solar Probe Observations during the First Encounter}
\author{L.-L. Zhao}
\affiliation{Center for Space Plasma and Aeronomic Research (CSPAR), The University of Alabama in Huntsville, Huntsville, AL 35805, USA}
\author{G. P. Zank}
\affiliation{Center for Space Plasma and Aeronomic Research (CSPAR), The University of Alabama in Huntsville, Huntsville, AL 35805, USA}
\affiliation{Department of Space Science, The University of Alabama in Huntsville, Huntsville, AL 35899, USA}
\author{L. Adhikari}
\affiliation{Center for Space Plasma and Aeronomic Research (CSPAR), The University of Alabama in Huntsville, Huntsville, AL 35805, USA}
\author{Q. Hu}
\affiliation{Center for Space Plasma and Aeronomic Research (CSPAR), The University of Alabama in Huntsville, Huntsville, AL 35805, USA}
\affiliation{Department of Space Science, The University of Alabama in Huntsville, Huntsville, AL 35899, USA}
\author{J. C. Kasper}
\affil{Department of Climate and Space Sciences and Engineering, University of Michigan, Ann Arbor, MI 48109, USA}
\affil{Smithsonian Astrophysical Observatory, Cambridge, MA 02138 USA}
\author{S. D. Bale}
\affil{Physics Department, University of California, Berkeley, CA 94720-7300, USA}
\affil{Space Sciences Laboratory, University of California, Berkeley, CA 94720-7450, USA}
\affil{The Blackett Laboratory, Imperial College London, London, SW7 2AZ, UK}
\affil{School of Physics and Astronomy, Queen Mary University of London, London E1 4NS, UK}
\author{K. E. Korreck}
\affil{Smithsonian Astrophysical Observatory, Cambridge, MA 02138 USA}
\author{A. W. Case}
\affil{Smithsonian Astrophysical Observatory, Cambridge, MA 02138 USA}
\author{M. Stevens}
\affil{Smithsonian Astrophysical Observatory, Cambridge, MA 02138 USA}
\author{J. W. Bonnell}
\affil{Space Sciences Laboratory, University of California, Berkeley, CA 94720-7450, USA}
\author{T. Dudok de Wit}
\affil{LPC2E, CNRS and University of Orl\'eans, Orl\'eans, France}
\author{K. Goetz}
\affil{School of Physics and Astronomy, University of Minnesota, Minneapolis, MN 55455, USA}
\author{P. R. Harvey}
\affil{Space Sciences Laboratory, University of California, Berkeley, CA 94720-7450, USA}
\author{R. J. MacDowall}
\affil{Solar System Exploration Division, NASA Goddard Space Flight Center, Greenbelt, MD 20771, USA}
\author{D. M. Malaspina}
\affil{Laboratory for Atmospheric and Space Physics, University of Colorado, Boulder, CO 80303, USA}
\author{M. Pulupa}
\affil{Space Sciences Laboratory, Univers ity of California, Berkeley, CA 94720-7450, USA}
\author{D. E. Larson}
\affil{Space Sciences Laboratory, University of California, Berkeley, CA 94720-7450, USA}
\author{R. Livi}
\affil{Space Sciences Laboratory, University of California, Berkeley, CA 94720-7450, USA}
\author{P. Whittlesey}
\affil{Space Sciences Laboratory, University of California, Berkeley, CA 94720-7450, USA}
\author{K. G. Klein}
\affil{Lunar and Planetary Laboratory, University of Arizona, Tucson, AZ 85721, USA}
\affil{Department of Planetary Sciences, University of Arizona, Tucson, AZ 85719, USA}

\begin{abstract}
	The Parker Solar Probe (PSP) observed an interplanetary coronal mass ejection (ICME) event during its first orbit around the sun, among many other events. This event is analyzed by applying a wavelet analysis technique to obtain the reduced magnetic helicity, cross helicity, and residual energy, the first two of which are magnetohydrodynamics (MHD) invariants.
Our results show that the ICME, as a large scale magnetic flux rope, possesses high magnetic helicity, very low cross helicity, and highly negative residual energy, thus pointing to a magnetic fluctuation dominated structure. Using the same technique, we also search for small-scale coherent magnetic flux rope structures during the period from 2018/10/22--2018/11/21, which are intrinsic to quasi-2D MHD turbulence in the solar wind. Multiple structures with duration between 8 and 300 minutes are identified from PSP in-situ spacecraft measurements. The location and scales of these structures are characterized by wavelet spectrograms of the normalized reduced magnetic helicity, normalized cross helicity and normalized residual energy. Transport theory suggests that these small-scale magnetic flux ropes may contribute to the acceleration of charged particles through magnetic reconnection processes, and the dissipation of these structures may be important for understanding the coronal heating processes.
\end{abstract}

\section{Introduction}\label{sec:introduction}

Some of the most important questions that Parker Solar Probe (PSP) intends to answer are how the solar corona is heated, and what processes accelerate suprathermal and energetic particles \citep[e.g.,][]{Fox16, Bale2016, Kasper2016, Bale2019, Kasper2019}. Magnetic reconnection and solar wind turbulence are two important phenomena that may be involved with both processes. Closely related to these processes are coherent structures such as small-scale magnetic flux ropes (SFRs). Magnetic flux ropes are helical magnetic field structures with approximately two-dimensional (2D) configuration and are also called magnetic islands or plasmoids. They have been frequently observed throughout the heliosphere. For example, \cite{Cart2010} identified and studied SFRs in the solar wind between 0.3 and 5.5 au using Helios, IMP 8, Wind, ACE, and Ulysses data. They found the occurrence rate of SFRs to be higher closer to the Sun and that SFRs generally lacked an expansion signature. Their observations support the view that SFRs are produced locally by magnetic reconnection across the heliospheric current sheet (HCS). \cite{Yu2014} studied a number of SFRs close to 1 au using Wind and STEREO data, and found that most of them were located in slow solar wind and did not have a significantly depressed proton temperature or plasma beta. 


The origin of SFRs is however not well understood. One view is that they are produced naturally via the cascade of quasi-2D turbulence. A common view of solar wind turbulence is that it consists of a majority 2D component and a minority slab component \citep{zank92, zank93, Zank2017}. Using an automatic Grad-Shafranov (GS) reconstruction technique, \cite{zheng18, Hu2018} identified tens of thousands of SFRs with scale sizes corresponding to the inertia range of turbulence using \emph{Wind} spacecraft data. The statistical analysis therein supports the idea that flux rope structures are representative of quasi-2D turbulence. Besides the above observations near 1 au, magnetic flux ropes are also observed beyond 1 au using \emph{Ulysses} measurements \citep[e.g.,][]{Chen2019, Zhao2019}. It is found that the properties of flux ropes pertinent to inertia-range turbulence persist at greater radial distance, and highly Alfv\'enic structures occur more frequently in high latitudes.

Some SFRs may originate from magnetic reconnection at the solar corona and be related to narrow CMEs/blobs observed in coronagraph white-light images. For example, \cite{Sanchez2017} find that coronal streamers can experience quasi-periodic bursts of activity with the simultaneous release of small transients or blobs. The signature of these transients includes helical magnetic fields and bidirectional streaming suprathermal electrons.   
In some cases, these blobs have been tracked all the way from the Sun to the Earth by heliospheric imagers, and were associated with SFRs in the solar wind \citep[e.g.,][]{Rouillard2010a, Rouillard2010b}.

Theories and simulations suggest that multiple interacting magnetic flux ropes can accelerate charged particles due to magnetic reconnection. The basic mechanisms include Fermi acceleration due to magnetic field line contraction \citep{Drake06}, and direct acceleration by anti-reconnection electric fields associated with the merging of magnetic islands or flux ropes \citep{Oka10}. Based on these basic mechanisms, \cite{Zank14} proposed a transport equation that describes particle acceleration in a ``sea'' of interacting magnetic islands and predicted power-law-like energy spectrum. From \emph{Ulysses} observations of an atypical energetic particle event \citep{Zhao2018, Zhao2019}, good agreement was found between the observed energetic proton intensity and the theoretical prediction based on the Zank et al. statistical transport model. A similar study was also presented for related observations at 1 au \citep{khaba15, khaba16, Adhikari2019}.

Another type of commonly observed structures in the solar wind are Alfv\'enic structures. In contrast to magnetic flux ropes that represent quasi-2D turbulence, Alfv\'enic structures are characteristic of slab turbulence \citep{Goldreich1995, Mont1995, Boldyrev2006, Mallet2015}. Alfv\'enic structures can be observationally similar to flux ropes as they also consist of a helical magnetic field. The difference is that flux ropes are nonpropagating structures that are convected with the plasma flow, whereas Alfv\'enic structures propagate at the local Alfv\'en speed along the mean magnetic field. Observations suggest that Alfv\'enic structures are more likely to be present in fast streams \citep[e.g.,][]{Bruno2013}. Slab turbulence/structures may also be related to particle energization due to stochastic heating \citep[e.g.,][]{Chandran10}.

The PSP mission allows us to further explore the region close to the sun. The objective of this paper is to identify and classify small-scale magnetic structures based on the unprecedented dataset returned from PSP during its first encounter. As discussed above, various observational techniques have been applied previously in studying these structures at 1 au, including the GS reconstruction. However, we have not applied the GS reconstruction to PSP data due to the Alfv\'enic nature of the solar wind that appears to be dominant in the observations. Alternatively, following previous studies of \cite{Telloni2012, Telloni2013}, we apply a wavelet analysis technique \citep{Torrence1998} to construct spectra of normalized reduced magnetic helicity, normalized cross helicity, and normalized residual energy based on time-series magnetic field and plasma parameters. These derived quantities provide additional and valuable information about the nature of the structures in the solar wind. This technique has the advantage that it can identify both Alfv\'enic structures and quasi-static magnetic flux ropes, and has been applied previously to observations at 1 au. Data from PSP/FIELDS \citep{Bale2016} and PSP/Solar Wind Electrons Alphas and Protons (SWEAP) \citep{Kasper2016} instruments during the period of one month near PSP's first perihelion (2018 October 22--2018 November 21) are analyzed in this paper. This is the first time that magnetic flux ropes are systematically identified within 0.3 au from the sun. 

The organization of this paper is as
follows. Section \ref{sec:technique} presents the basic procedures of our analysis technique. Section \ref{sec:results} shows the examples of the identified flux ropes, including the large-scale ICME event. A statistical analysis of normalized residual energy and normalized cross helicity for structures with high magnetic helicity is also presented.
The last section provides a summary and conclusion.

\section{Analysis technique}\label{sec:technique}

It is generally accepted that the most plausible magnetic configuration of magnetic flux ropes consists of helical field lines winding around a central axis \citep[e.g.,][]{Burlaga1988}. They are thus expected to possess a high value of magnetic helicity, which is a conserved quantity of the ideal magnetohydrodynamic (MHD) equation and characterizes the knottedness of magnetic field lines \citep[e.g.,][]{Matth1982}. 
To distinguish Alfv\'enic fluctuations from a flux rope structure in quasi-static equilibrium, two other important parameters, namely the cross helicity and residual energy, are also employed to determine the presence of Alfv\'en waves. Although clear signatures of Alfv\'en waves had been observed coinciding with magnetic clouds and SFRs \citep{Marsch2009, Gosling2010, Yao2010, Gershman2017}, such events are very rare \citep[see, e.g.,][]{Gosling2010}. 

We follow the method of \cite{Telloni2012} and use a Morlet wavelet analysis to study the signatures of these three parameters via the observed magnetic field and plasma parameters. Fluctuating and mean magnetic and velocity fields can be separated as $\underline{\textbf B} = \textbf B_0 +\textbf b$; $\underline{\textbf U} = \textbf U_0 +\textbf u$. Here, $\textbf B_0$ is the mean magnetic field, $\textbf U_0$ is the mean velocity field, $\textbf b$ represents the fluctuating magnetic field, and $\textbf u$ represents the fluctuating velocity field. The mean magnetic field is $\langle\underline{\textbf B}\rangle = \textbf B_0$ with $\langle \textbf b\rangle=0$, and similarly for the velocity field. 

The strict definition of magnetic helicity density is the dot product of the magnetic vector potential and the magnetic field, which depends on the spatial properties of the magnetic field topology, and thus cannot be directly evaluated from single spacecraft measurements. However, \cite{Matth1982} described a reduced form of magnetic helicity that can be estimated with measurements from a single spacecraft based on the magnetic power spectrum. We then perform the Morlet-wavelet transforms \citep{Torrence1998} on each component of the fluctuating magnetic field $b_R$, $b_T$, $b_N$ to compute the magnetic power spectrum tensor. According to \cite{Matth1982}, the normalized reduced magnetic helicity can be estimated by 
\begin{equation}\label{sigmam}
  \sigma_m(\nu,t) = \frac{2 \operatorname{Im}[W_T^*(\nu,t) \cdot W_N(\nu,t)]}{|W_R(\nu,t)|^2 + |W_T(\nu,t)|^2 + |W_N(\nu,t)|^2},
\end{equation}
where $\nu$ is the frequency associated with the Wavelet function and the sampling period of the measured magnetic field in the RTN coordinate system. Here, we average the magnetic field data from PSP/FIELDS measurements down to 30-second cadence to comply with the resolution of plasma data. The spectra $W_R(\nu,t)$, $W_T(\nu,t)$ and $W_N(\nu,t)$ are the wavelet transforms of time series of $b_R$, $b_T$ and $b_N$, respectively, and $W^*_T(\nu,t)$ is the conjugate of $W_T(\nu,t)$. From the resulting spectrogram of the magnetic helicity $\sigma_m$, one can determine both the magnitude and the handedness (chirality) of underlying fluctuations at a specific scale. 
A positive value of $\sigma_m$ corresponds to right-handed chirality and a negative value to left-handed chirality. 

The normalized cross helicity $\sigma_c$ and residual energy $\sigma_r$ are usually calculated from the Els\"asser variables $\textbf z^\pm= \textbf u \pm \tilde{\textbf b}$ with $\tilde{\textbf b} = \textbf b/\sqrt{4\pi n_p m_p}$, $n_p$ the proton density, and $m_p$ proton mass \citep[e.g.,][]{Zank2012}:
\begin{equation}\label{sigmac1}
\sigma_c = \frac{\langle z^{+2} \rangle - \langle z^{-2}\rangle}{\langle z^{+2} \rangle + \langle z^{-2}\rangle} = \frac{2 \langle \textbf u \cdot \tilde{\textbf b}\rangle}{\langle u^2 \rangle + \langle \tilde{b}^2 \rangle},
\end{equation}
and
\begin{equation}\label{sigmad1}
\sigma_r = \frac{2 \langle \textbf z^{+} \cdot \textbf z^- \rangle}{\langle z^{+2} \rangle  + \langle z^{-2} \rangle} = \frac{\langle u^2 \rangle - \langle \tilde{b}^2 \rangle}{\langle u^2 \rangle + \langle \tilde{b}^2 \rangle},
\end{equation}
where $\textbf z^+$ ($\textbf z^-$) represents the forward (backward) propagating modes with respect to the magnetic field orientation, and $\langle{z^+}^2\rangle$ and $\langle{z^-}^2\rangle$ respectively represent the energy density in forward and backward 
propagating modes. The absolute values of $\sigma_m$, $\sigma_c$ and $\sigma_r$ are no more than 1. 
The magnitude of $\sigma_c$ indicates the alignment between $\textbf b$ and $\textbf u$ provided that the magnitude $\textbf u$ is significant. Unidirectional Alfv\'en waves usually have a high value of $|\sigma_c|$ (close to 1). More energy resides in forward propagating Alfv\'en wave modes if $\sigma_c > 0$ and in backward propagating modes (with respect to the orientation of the mean magnetic field) if $\sigma_c <0$. The normalized residual energy $\sigma_r$ represents the energy difference between the fluctuating kinetic and magnetic energies. Magnetic fluctuating energy dominates when $\sigma_r<0$, and kinetic fluctuating energy dominates when $\sigma_r>0$. Alfv\'en waves usually have a typical $\sigma_r$ close to zero. 

To obtain the corresponding spectrograms, we further perform the wavelet transform $\mathcal{W}$ on the three components of the Els\"asser variables $ z_R^\pm$, $z_T^\pm$, and $z_N^\pm$. The spectrograms of normalized residual energy $\sigma_r$ and cross helicity $\sigma_c$ can be rewritten in both the frequency and time domains as
\begin{equation}\label{sigmar}
\sigma_r(\nu, t) = \frac{2 \operatorname{Re}[\mathcal{W}^*(z_R^+) \cdot \mathcal{W}(z_R^-) + \mathcal{W}^*(z_T^+) \cdot \mathcal{W}(z_T^-) + \mathcal{W}^*(z_N^+) \cdot \mathcal{W}(z_N^-)]}{W^{+}(\nu,t) + W^{-}(\nu, t)},
\end{equation}
 and
\begin{equation}\label{sigmacc}
\sigma_c(\nu, t)= \frac{W^{+}(\nu,t) - W^{-}(\nu, t)}{W^{+}(\nu,t) + W^{-}(\nu, t)},
\end{equation}
where $W^+(\nu, t)$ and $W^-(\nu, t)$ represents the wavelet power spectrum in $\textbf z^+$ and $\textbf z^-$ modes, respectively, i.e., $W^+(\nu, t)=|\mathcal{W}(z_R^+)|^2 + |\mathcal{W}(z_T^+)|^2 + |\mathcal{W}(z_N^+)|^2$ and $W^-(\nu, t)=|\mathcal{W}(z_R^-)|^2+ |\mathcal{W}(z_T^-)|^2 + |\mathcal{W}(z_N^-)|^2$.

\section{Results}\label{sec:results}
\begin{figure}[htbp]
\centering
\includegraphics[width=1.0\linewidth]{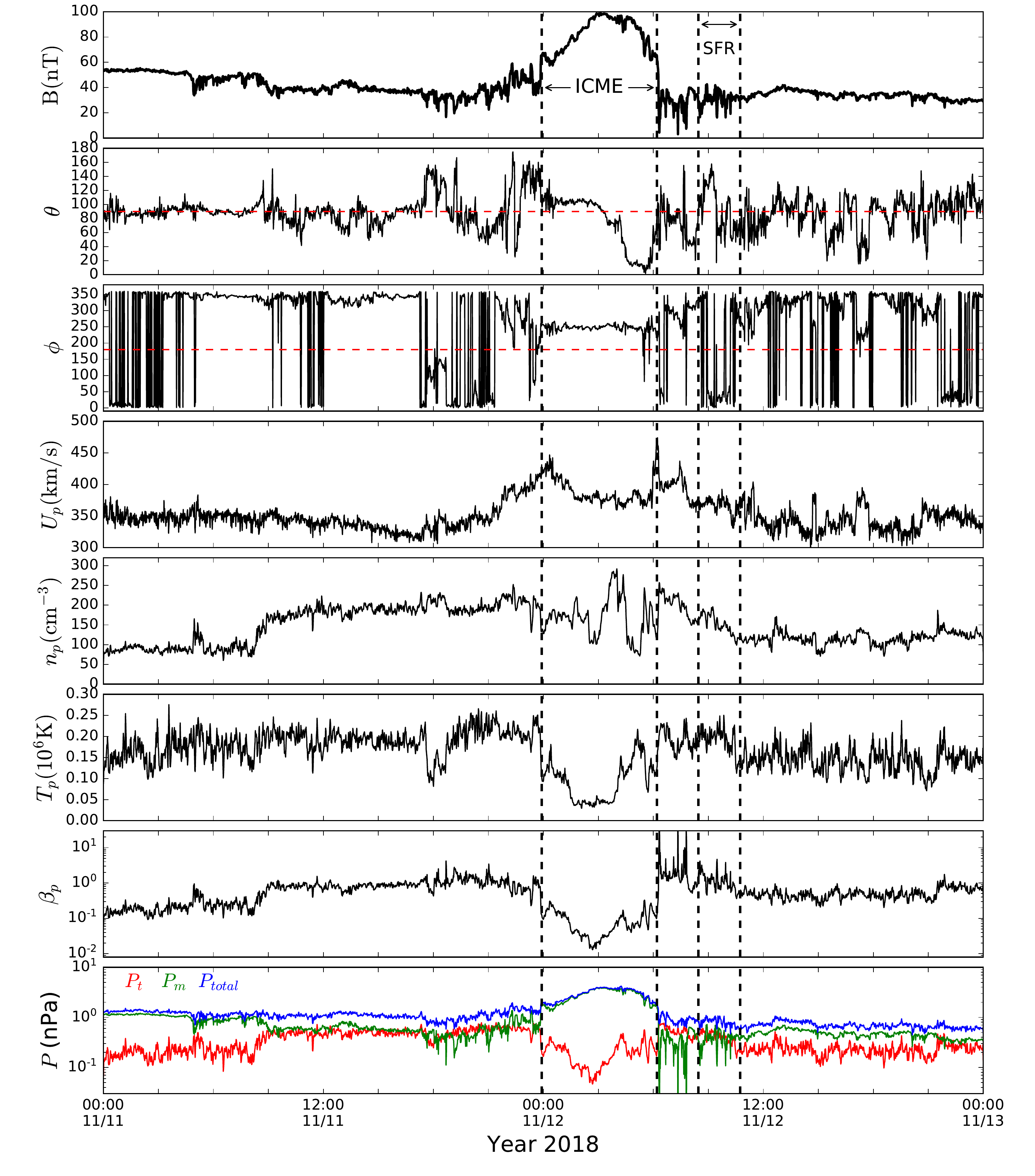} 
\caption{The PSP in situ observations from 2018 November 11, 00:00 UT to 2018 November 13, 00:00 UT. The panels from top to bottom show, respectively, the magnetic field magnitude ($B$), the elevation ($\theta$) and azimuthal ($\phi$) angles of the magnetic field direction in the RTN coordinate system, solar wind speed ($U_p$), proton number density ($n_p$), proton temperature ($T_p$), proton beta ($\beta_p$), and the thermal ($P_t$), magnetic ($P_m$), and total pressure $P_{total}$. The dashed vertical lines mark the ICME interval and SFR interval.}\label{fig:CMEparameter}
\end{figure}

The PSP observed an ICME event at $\sim$0.26 au on 2018 November 12 during its first orbit around the sun \citep[e.g.,][]{Giacalone2019}.
Figure \ref{fig:CMEparameter} displays a two-day time-series plot of the varying magnetic field and plasma parameters from 2018 November 11, 00:00 UT to 2018 November 13, 00:00 UT measured by the PSP/FIELDS and PSP/SWEAP instruments. The panels from top to bottom show, respectively, the magnetic field strength $B$ with 1-minute cadence, the elevation ($\theta$) and azimuthal ($\phi$) angles of the magnetic field direction in the RTN coordinate system, the 1-minute averaged flow speed ($U_p$), proton density ($n_p$), proton temperature ($T_p$), proton beta ($\beta_p$), proton thermal pressure ($P_t$), magnetic pressure ($P_m$), and the total pressure $P_{total} = P_t+P_m$. The proton beta, thermal, magnetic, and total pressure are plotted in log scale. 

During this period, the magnetic field is mainly in the $R$ direction, since the associated directional angles are $\theta \simeq 90^{\circ}$, $\phi \simeq 360^{\circ}$ or $0^{\circ}$ for most of the time. The magnetic field magnitude is around 60 nT, and solar wind speed is about 350 km/s until $\sim$21:00 UT on November 11, when the flow speed begins to increase. After that, the flow velocity gradually increases to $\sim$450 km/s, and the fluctuations in the magnetic field begin to become more prominent, with larger amplitude and frequent directional changes. During the period from November 11, 23:55 UT to November 12, 06:12 UT, PSP observed an ICME event at $\sim$0.26 AU, as indicated by the interval between the first two dashed vertical lines. The ICME lasts for about 6 hours. The dominant ICME signatures \citep[e.g.,][]{Kilpua2017} in this event are abnormally low proton temperature and $\beta_p$, enhanced magnetic field strength and total pressure, and a large-scale smooth rotation of the magnetic field vector that signifies its flux-rope geometry. Within the ICME interval, the magnetic field strength reaches a maximum value of $\sim$100 nT, the proton temperature $T_p$ reaches below $10^5$ K, the solar wind flow velocity varies between 370 and 450 km/s, and the proton plasma beta reaches a minimum value of $\sim$0.01 due to the large increase in magnetic pressure and the decrease in thermal pressure. This ICME event may be short in duration, but it is still at the lower end of large-scale heliospheric transients and could indeed originate from the Sun, given also the close proximity of PSP to the Sun. There is a SFR structure after this ICME event, as indicated by the last two dashed vertical lines. The approximate time interval of this SFR starts from 08:27 UT to 10:44 UT, and lasts for 137 minutes. The dominant signature of this SFR is the rotation of the magnetic field direction. We discuss other features of small-scale flux rope structures in detail below.

\begin{figure}[htbp]
\centering
\includegraphics[width=1.0\linewidth]{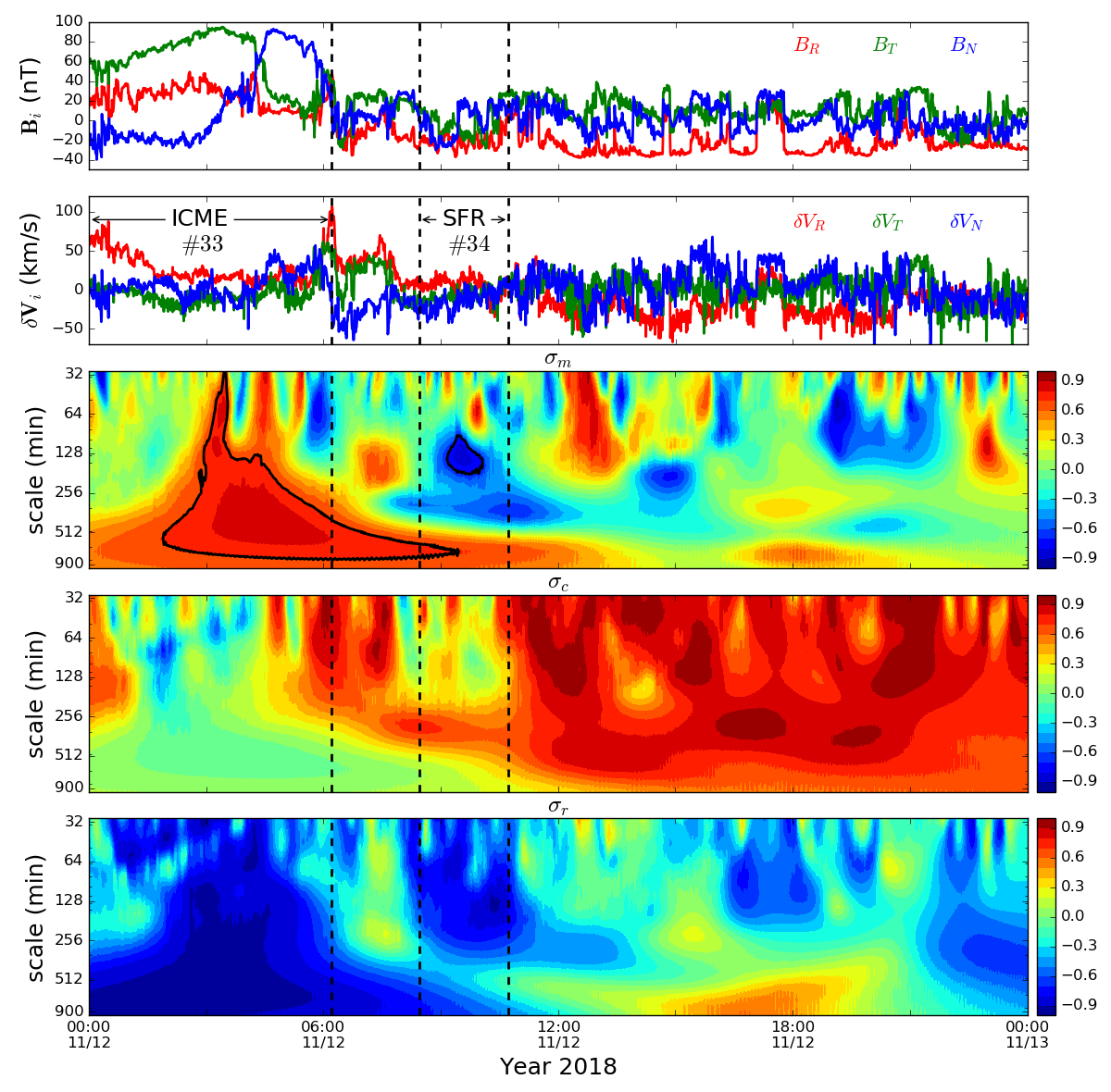} 
	\caption{The top two panels show time profiles of the magnetic field vector and the plasma velocity fluctuations with an average time resolution of 30 seconds on 2018 November 12 measured by PSP/FIELDS and PSP/SWEAP instruments, respectively. The ICME and SFR (\#33 and \#34 in Table 1, respectively) are identified by the vertical dashed lines the same interval as marked in Figure \ref{fig:CMEparameter}. The bottom three panels show spectrograms of the normalized reduced magnetic helicity $\sigma_m$, normalized cross helicity $\sigma_c$, and normalized residual energy $\sigma_r$, using a Morlet wavelet analysis. Contour lines are drawn at levels $|\sigma_m|=0.7$.}\label{fig:CMEheli}
\end{figure}

We apply the method described in Section \ref{sec:technique} to first analyze this ICME event, which is considered as a large-scale magnetic flux rope or typically called a magnetic cloud. The top two panels of Figure \ref{fig:CMEheli} show the time profiles of the $R$, $T$, and $N$ components of the magnetic field and the plasma velocity fluctuations measured by PSP/FIELDS and PSP/SWEAP instruments, respectively, with a uniform time resolution of 30 seconds on 2018 November 12. The ICME starts from $\sim$00:00 UT and ends at $\sim$06:12 UT (denoted by the vertical dashed lines in the figure) with a smooth rotation of the $B_N$ component.
Within the ICME, the plasma velocity fluctuations are extremely low with $\delta V_i \sim 0$. After the crossing of the ICME, both the plasma velocity and the magnetic field fluctuate rapidly and positively correlate with each other, indicating that the Alfv\'enic fluctuations are generated in the region downstream of the ICME. The following three panels of Figure \ref{fig:CMEheli} display the spectrograms of the normalized reduced magnetic helicity $\sigma_m$, normalized cross helicity $\sigma_c$, and normalized residual energy $\sigma_r$, obtained by applying Equations \eqref{sigmam}, \eqref{sigmar}, and \eqref{sigmacc}, respectively, and using the Morlet wavelet. The wavelet scales are chosen to be between $\sim$30 minutes and $\sim$16 hours in these plots.
The contour lines in the panel for $\sigma_m$ enclose high magnetic helicity regions with $|\sigma_m| \ge 0.7$. The ICME is clearly identified in the spectrogram of the normalized reduced magnetic helicity $\sigma_m$ as a right-handed magnetic helical structure with a high value of $|\sigma_m|$. The averaged $\sigma_m$ over the ICME region bounded by the black contour line is 0.79, the averaged $\sigma_c$ is 0.01, and the averaged $\sigma_r$ is -0.89. The extremely low value of $\sigma_c$ indicates that there is no Alfv\'en wave fluctuations in this ICME event. The normalized residual energy $\sigma_r$ is highly negative, indicating that the magnetic fluctuation energy dominates the ICME interval. Overall the velocity fluctuations are negligible. On the contrary, the plasma downstream of the ICME behaves as a typically outwardly propagating Alfv\'en waves, characterized by $\sigma_c \sim 1$ most of the time, which is coincident with the characteristics shown in the time profiles of the magnetic field vector and plasma velocity fluctuations.

As a large-scale magnetic flux rope structure, the ICME on 2018 November 12 shows high magnetic helicity, near zero cross helicity, and high negative residual energy, thus indicating that flux ropes can be identified according to the characteristics of the spectral features in magnetic helicity, cross helicity, and residual energy. 
In fact, the rotation of magnetic field components in flux rope events can result in high magnetic helicity, and additionally low values of cross helicity and non-zero negative residual energy can be further used to exclude the Alfv\'enic structures. 
As an example, we identify a SFR structure bounded by the contour lines at $\sim$09:30 UT just after the ICME, which has much smaller plasma velocity fluctuations compared to the surrounding medium as shown in the second panel of Figure \ref{fig:CMEheli}. This flux rope is also clearly characterized by a large $|\sigma_m|$, a small $|\sigma_c|$, and a highly negative $\sigma_r$. The averaged $\sigma_m$ over the bounded region is -0.77, the averaged $\sigma_c$ is 0.24, and the averaged $\sigma_r$ is -0.75. The spectrograms in Figure \ref{fig:CMEheli} suggest that it is a left-handed helical magnetic structure with a scale of about 130 minutes. 

We now apply the wavelet technique to the data set for the time period from 2018 October 22 to 2018 November 21, which is a month surrounding PSP's first perihelion at around 35 solar radius on 2018 November 6.
Based on the above discussions, we set the following threshold conditions for the detection of SFRs: (\romannumeral1) the normalized reduced magnetic helicity $|\sigma_m| > 0.7$; (\romannumeral2) the normalized cross helicity $|\sigma_c| < 0.4 $; and (\romannumeral3) the normalized residual energy $\sigma_r < -0.5$. An event candidate will be identified when all three criteria are met simultaneously. The corresponding scales for identified events will also be recorded.
One may argue that Alfv\'en waves, with a large value of $|\sigma_c|$ and a close-to-zero $\sigma_r$, can also exist in flux ropes.
However, since we cannot give criteria to distinguish between the stand-alone Alfv\'en waves and Alfv\'en waves co-existing within a magnetic flux rope, we elect to follow \cite{Cart2010, Hu2018} and exclude all structures with clear Alfv\'enic fluctuations as possible flux ropes. 

Following the above procedure, a total of 40 structures are identified and they are listed in Table \ref{tab:ropes}. Note that the ICME event corresponds to structure \# 33 in the table.
For each of the detected structures, we calculate its central time $t$ and scale $s$ using the ``center of mass'' of the corresponding contour on the spectrogram, weighted by the normalized magnetic helicity:
\begin{equation}
  t = \frac{\sum_{i} \sigma_{mi} t_i}{\sum_{i} \sigma_{mi}}; \qquad
  s = \frac{\sum_{i} \sigma_{mi} s_i}{\sum_{i} \sigma_{mi}}.
\end{equation}
Here, the summation is done over the region enclosed by each contour on the spectrogram, and the subscript $i$ refers to an individual point inside the contour with corresponding time $t_i$ and scale $s_i$, respectively. Similarly, the averaged $\sigma_m$, $\sigma_c$, and $\sigma_r$ are also calculated for each structure. These parameters for the identified flux ropes are listed in Table 1. Also listed in Table 1 is the average solar wind speed and proton beta for each SFR event. We find that almost all the identified SFRs lie in slow solar wind and possess a wide range of proton beta. However, we remark that this is only a feature of the most probable SFRs during PSP's first encounter, and that it may not account for all the SFRs in the inner heliosphere. Those days for which four or more SFRs were detected (10/24, 10/28, 10/29, 11/11) may be related to HCS crossings. \cite{Szabo2019} have identified HCS crossings observed by PSP, which reveals a more complex structure than at 1 AU. Numerous discontinuities and possible magnetic reconnection signatures have been detected within HCS crossing regions. The connection between identified SFRs and HCS crossing will be the subject of further study.

\begin{scriptsize}
\begin{longtable}{cccccccc}
\caption{List of identified magnetic flux ropes from 2018/10/22 to 2018/11/21}\\
\hline \noalign {\smallskip}
No. &  Central time & Scale & $<\sigma_m>$ &  $<\sigma_c>$  & $<\sigma_r>$ & $<V_{sw}>$ & $<\beta_p>$\\
    &      (UT)        & (min)    &              &                &    & (km/s) &     \\
\hline \noalign {\smallskip}
1  & 09:15 10/22 & 8 & -0.78 & 0.30 & -0.52 & 281 & 1.21\\
2 &14:22 10/23& 11&0.74&0.19&-0.86 & 376 & 0.61\\ 
3  &07:58 10/24 &12 &-0.76 &0.08&-0.68 &409 & 0.01\\ 
4 &08:51 10/24 &8&-0.74& 0.19 &-0.87 & 359 & 0.01\\
5 &17:24 10/24&75&0.75&0.31& -0.74 & 397&0.02 \\
6  &23:49 10/24& 16&-0.78&0.18&-0.70 & 383 & 0.01\\
7  &10:37 10/26& 15& 0.75 &0.08 &-0.64 & 295 & 1.02\\
8  &11:02 10/26& 22&-0.78 & -0.15 &-0.64 & 292 & 1.38\\
9  &12:17 10/26 &17 &0.74 &0.36 &-0.56 & 287 & 0.76 \\
10  &05:17 10/27 &67 &-0.75 &0.18 &-0.85 & 292 & 0.74\\
11  &03:12 10/28& 9&0.79 &-0.16 &-0.56 & 302 & 2.20\\
12  &04:30 10/28 & 71&-0.74 &0.20 &-0.88 & 293 & 1.58\\
13  & 05:11 10/28& 31 &0.78  &0.38 &-0.65 & 291 & 0.71\\
14  & 05:46 10/28& 59 &-0.75 &-0.01 &-0.69 & 290 & 1.1\\
15  &23:23 10/28 & 41&0.71 &-0.24 &-0.90 & 265 & 0.33\\
16  &04:47 10/29 &19&0.78 &-0.14 &-0.68 & 280 & 1.00 \\
17  &13:09 10/29 & 35 &0.72 &-0.26 &-0.50 & 293 & 3.27\\
18   &17:31 10/29 & 14&0.71 &0.23 &-0.68 & 325 & 0.60\\
19  &17:51 10/29 & 9 &0.77 &0.26 &-0.74 & 323 & 0.66\\
20  &18:26 10/29 &18 &-0.77 &-0.07 &-0.50 & 335 & 0.60 \\
21  &19:59 10/29 & 43 &-0.82 &0.19 &-0.75 & 330 & 1.05 \\
22  &04:16 10/30 & 44 &0.82 &0.10 &-0.89 & 333 & 0.37 \\
23  &00:29 10/31 & 15 &0.81 &-0.14 &-0.63 & 309 & 0.18 \\
24  &03:48 10/31 & 15 &-0.74 &0.27 &-0.70 & 348 & 0.40 \\
25  &07:42 10/31 & 65 &0.80 & 0.38 &-0.60 & 332 & 0.08 \\
26  & 21:17 11/04 & 25 &-0.78 &0.36 &-0.76  & 303 & 0.18 \\
27  &12:30 11/11 & 40 & 0.79 &0.19 & -0.59  & 344 & 0.87 \\
28   &19:58 11/11 & 15 & 0.73 & -0.08 & -0.76 & 342 & 1.45 \\
29   &20:25 11/11 & 48 & -0.72 & 0.31 & -0.79 & 348 & 1.21 \\
30  &22:19 11/11 & 31 &-0.75 & 0.17 & -0.93 & 390 & 0.77 \\
31  &22:54 11/11 & 18 & -0.72 & -0.15 & -0.79 & 392 & 0.63 \\
32   &01:52 11/12 & 16 &0.72 &0.15 &-0.64 & 378 & 0.04 \\
33  &04:00 11/12 & 264 & 0.79 & 0.01 & -0.89 & 380 & 0.07 \\ 
34   &09:34 11/12 & 137 & -0.77 &  0.24 & -0.75 & 367 & 1.3\\ 
35  &09:35 11/13 & 10 & 0.73 & 0.20 & -0.76 & 353 & 0.49\\
36  &10:41 11/13 & 116 &0.74 & 0.14 &-0.82 & 352 & 0.88\\    
37   &23:35 11/13 & 29 & -0.78 & 0.35 &-0.81 & 325 & 0.90\\
38   &02:56 11/14 &102 & -0.80 &  -0.00 & -0.75 & 321 & 0.67\\
39  &17:47 11/14 & 47 & -0.73 & 0.19 & -0.78 & 391 & 0.70\\
40   &00:55 11/15 & 36 &  0.73 & -0.19 & -0.60 & 520 & 0.31\\
\label{tab:ropes}
\end{longtable}

\end{scriptsize}

\begin{figure}[!htbp]
\centering
\includegraphics[width=1.0\linewidth]{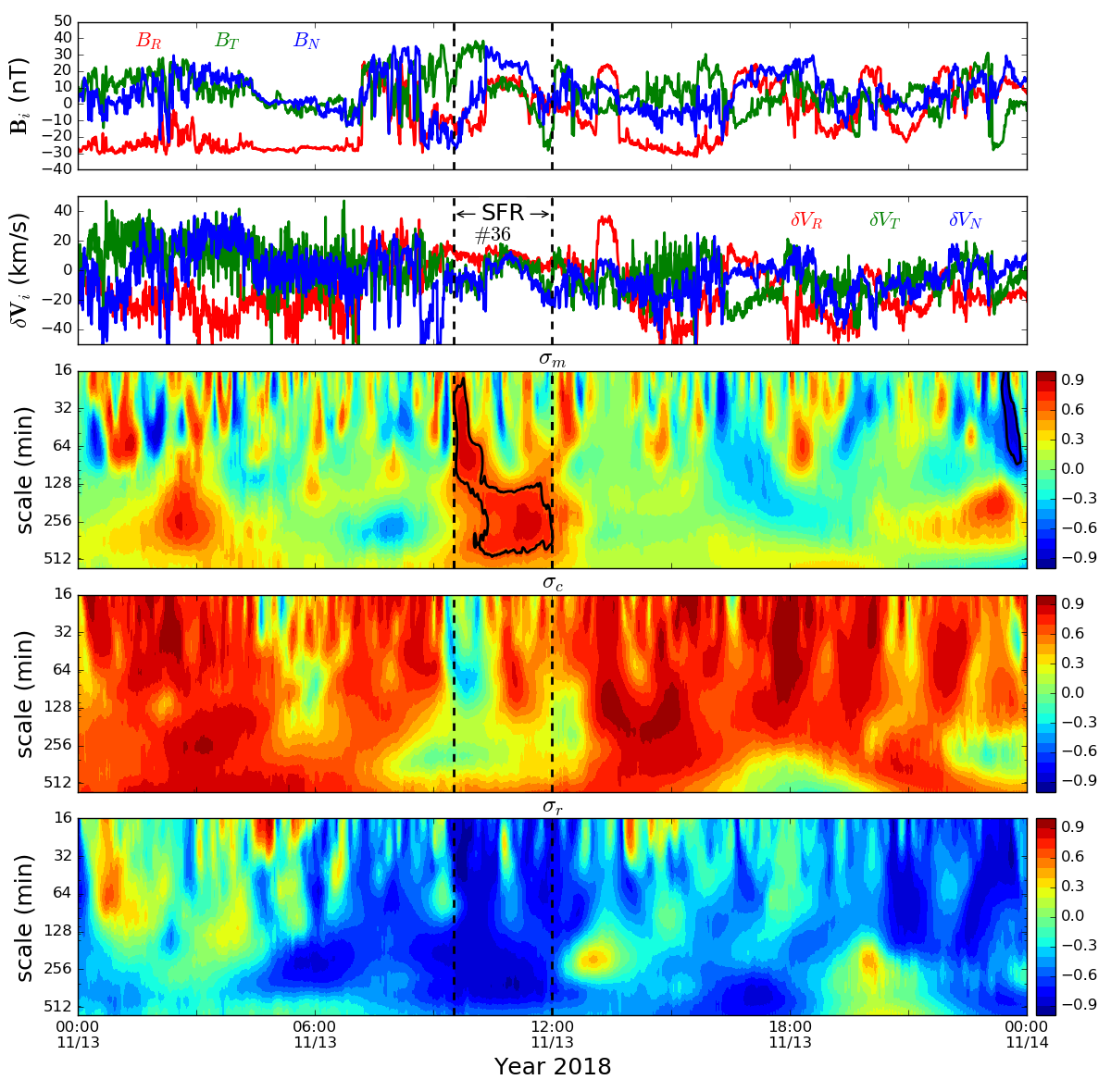}\\ 
\caption{Same as Figure \ref{fig:CMEheli}, but for the flux rope event on 2018 November 13 (\#36 in Table \ref{tab:ropes}).}\label{fig:1113}
\end{figure}

Figure \ref{fig:1113} shows an example of the identified small-scale magnetic flux rope occurring on 2018 November 13, corresponding to \# 36 in Table \ref{tab:ropes}. The vertical dashed lines in the figure delimit the time period of the identified flux rope with a smooth rotation of the magnetic field vector over about 150 minutes. The plasma velocity fluctuation level is much lower compared to the surrounding plasma. The wavelet spectrograms of $\sigma_m$, $\sigma_c$ and $\sigma_r$ exhibit typical values for a flux rope structure, which indicates a right-handed helical structure characterized by high magnetic helicty with $<\sigma_m>= 0.74$, low cross helicity with $<\sigma_c> = 0.14$, and high negative residual energy with $<\sigma_r>=-0.82$. There seems to exist two peaks inside the contour on the spectrogram for $\sigma_m$. This is due to there being two consecutive rotations in the magnetic field vectors. The first peak of $\sigma_m$ in the scale domain is located at about 50 minutes, and the corresponding time is at $\sim$09:40 UT. The second peak occurs at $\sim$ 11:00 UT with a scale of $\sim$220 minutes. This flux rope seems to be strictly limited to the time interval identified, since its magnetic and cross-helicity values are indeed well isolated from the surrounding medium. The $\sigma_m$ outside of the flux rope interval is near zero with $\sigma_c$ close to 1. The upstream and downstream regions appear to be quite Alfv\'enic. This may indicate that this flux rope is embedded in an Alfv\'enic stream, probably generated locally via solar wind turbulent reconnection.

\begin{figure}[!htbp]
\centering
\includegraphics[width=0.5\linewidth]{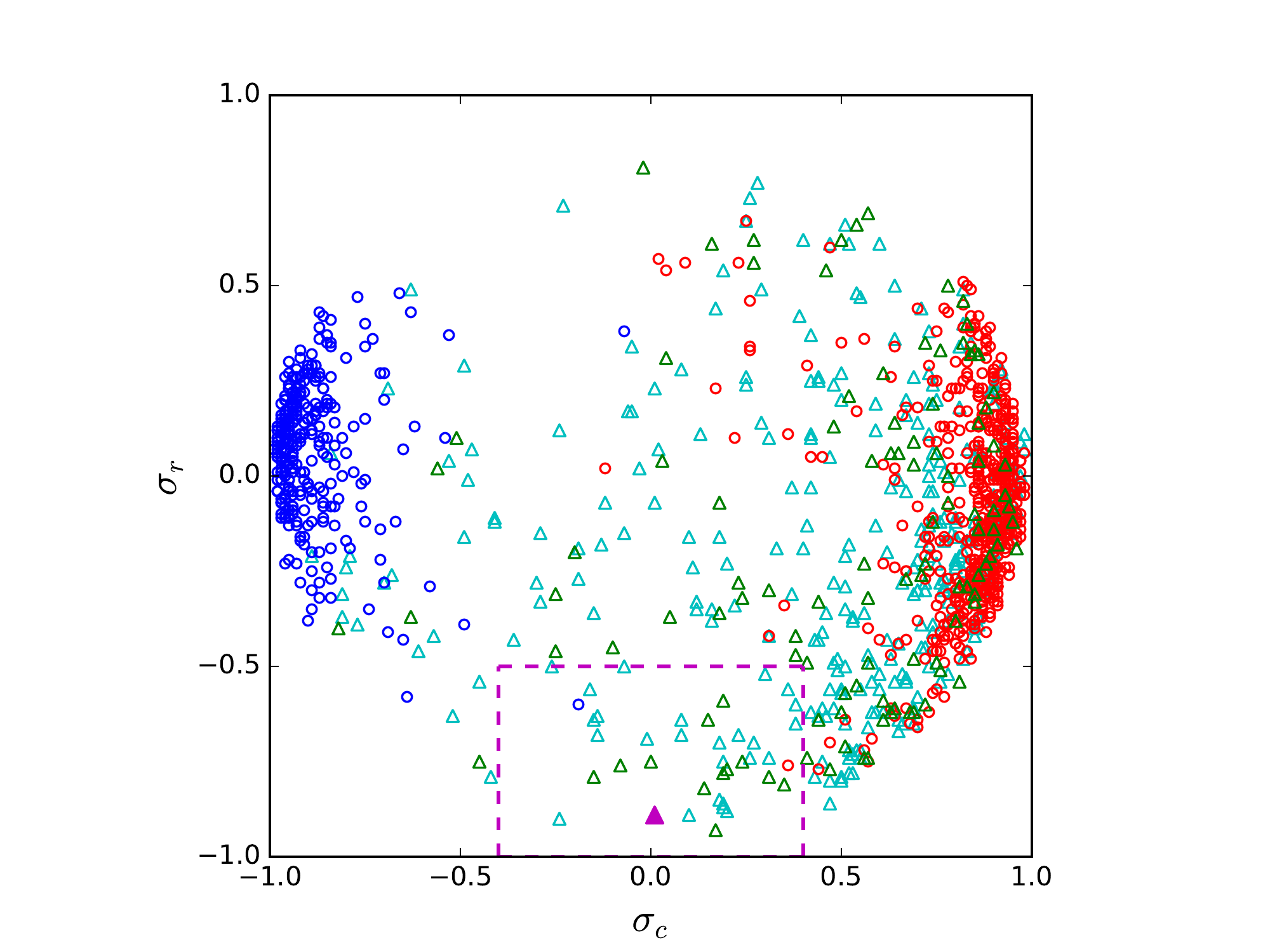}\\
\includegraphics[width=0.55\linewidth]{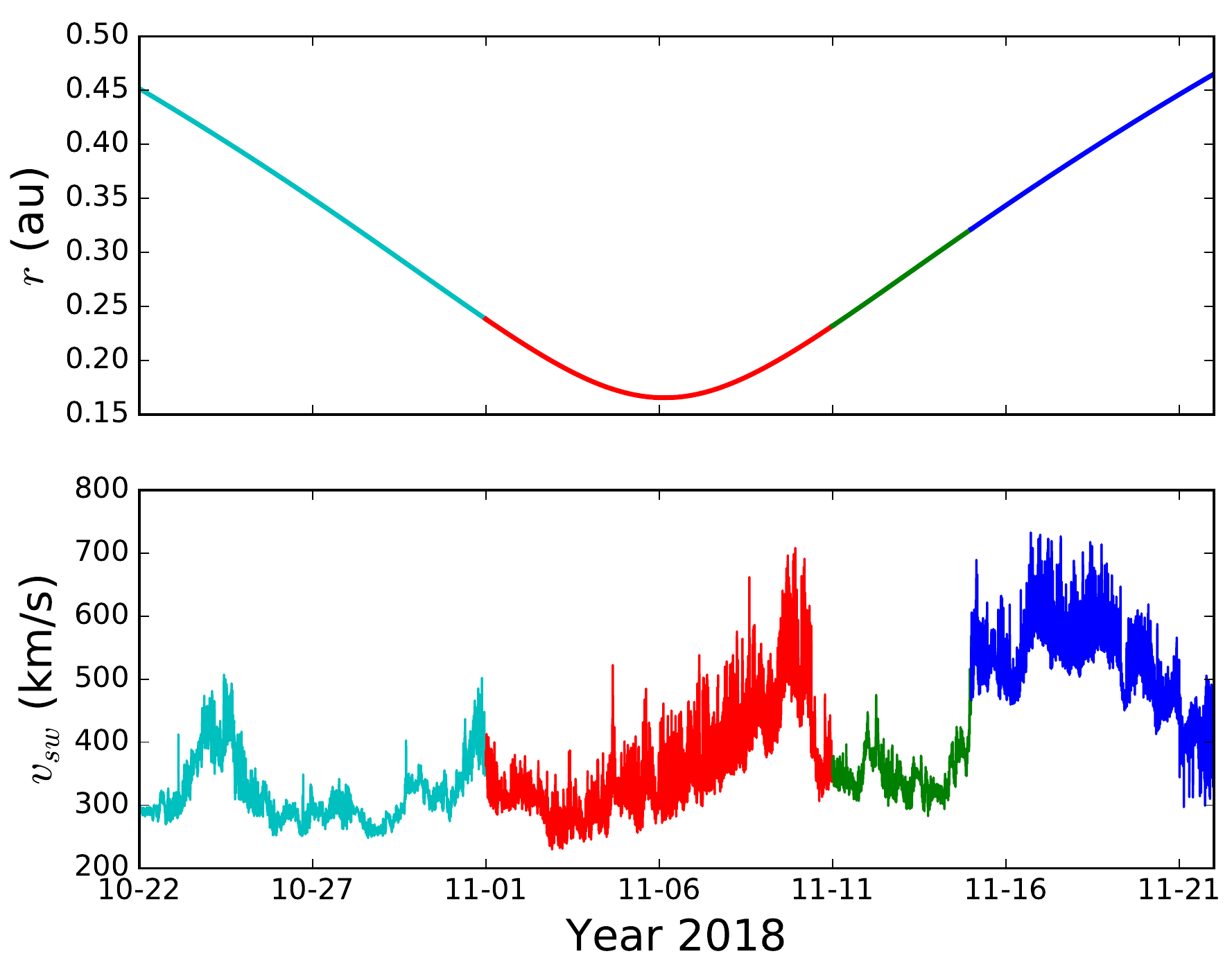}
\caption{The top panel shows the normalized residual energy $\sigma_r$ versus normalized cross helicity $\sigma_c$ for all structures with $|\sigma_m|>0.7$ during the period from 2018 October 22 to 2018 November 21. The structures are grouped into four categories corresponding to the four time periods, and are represented by cyan triangles for the period (a) 2018 October 22--October 31, red circles for the period (b) 2018 November 1--November 10, green triangles for the period (c) 2018 November 11--November 14, and blue circles for the period (d) 2018 November 15--November 21. The rectangular box represents the region that likely contains flux rope structures with small cross helicity $|\sigma_c| < 0.4$ and negative residual energy $\sigma_r < -0.5$. The bottom two panels show the corresponding radial distance from the PSP to the sun and 10-minute moving averaged solar wind speed, for the same time period, color-coded according to the groups in the top panel.} \label{fig:psp}
\end{figure}

To obtain a collective view on structures with enhanced $\sigma_m$ including magnetic flux ropes, we identify all events with $|\sigma_m| > 0.7$ only. As a result, 1245 structures are found in the spectrograms during the same time period as Table \ref{tab:ropes}.
Note that we have filtered out structures that have scales smaller than 8 minutes and larger than 300 minutes. Structures with scales less than 8 minutes may be contaminated by discontinuities, such as current sheets and magnetic switchbacks \citep{McManus2019}, and are not the focus of this study. The very large structures may not be truly reliable since they usually fall outside the cone of influence (COI) of the wavelet spectra \citep{Torrence1998}.
For each structure, we again calculate its averaged $\sigma_m$, $\sigma_c$, and $\sigma_r$. In the top panel of Figure \ref{fig:psp}, we plot the normalized residual energy $\sigma_r$ versus normalized cross helicity $\sigma_c$ for all these structures.
To put it in context, we plot the radial distance from the PSP to the sun and 10-minute moving averaged solar wind speed during the corresponding time period in the bottom two panels of Figure \ref{fig:psp}.
Based on the radial distance and the solar wind speed, we divide the month into 4 segments: (a) 2018 October 22--October 31; (b) 2018 November 1--November 10; (c) 2018 November 11--November 14; and (d) 2018 November 15--November 21. The four periods are colored in cyan, red, green, and blue in Figure \ref{fig:psp}, respectively. The figure suggests that the periods (a) and (c) are mostly dominated by slow solar wind, and the period (d) is dominated by fast wind. During the period (b), PSP is near the perihelion, and the solar wind speed exhibits large fluctuations.
Corresponding to the four time periods, the identified structures are also grouped into four subsets, and they are represented in the top panel of Figure \ref{fig:psp} by cyan triangles for the period (a), red circles for the period (b), green triangles for the period (c), and blue circles for the period (d).
The rectangular box in Figure \ref{fig:psp} represents the region that most likely contains the identified flux rope structures, corresponding to the criteria that we set previously ($|\sigma_c| < 0.4$ and $\sigma_r < -0.5$). The ICME event identified in Figure \ref{fig:CMEparameter} satisfies our criteria and is therefore included in our flux rope list (event \# 33 in Table \ref{tab:ropes}), which is highlighted as a magenta triangle in the figure.
Figure \ref{fig:psp} shows that almost all flux rope candidates are observed within periods (a) and (c). This strongly suggests that the magnetic flux ropes are associated with the slow solar wind during this PSP orbit period, which is in good agreement with previous statistical studies \citep[e.g.,][]{Yu2014}. On the other hand, structures in periods (b) and (d) are predominantly Alfv\'enic, as illustrated by the large $|\sigma_c|$ and small $|\sigma_r|$ values. Another feature is that periods (b) and (d) are characterized by opposite signs of $\sigma_c$, indicating oppositely propagating wave modes in these two periods.
We do not rule out the possibility that the structures outside the box can be flux ropes. For example, structures with cross helicity and residual energy in the range $|\sigma_c| < 0.4$ and $-0.5 < \sigma_r < 0$ may also be magnetic flux ropes if the remaining flow is much smaller (close to zero), although they do not satisfy the criteria that we set. Other types of structures such as flow vortices with $\sigma_r>0$ may also be present in Figure \ref{fig:psp}. However, these events are not the focus of this study.

\begin{figure}[htbp]
\centering
\includegraphics[width=0.55\linewidth]{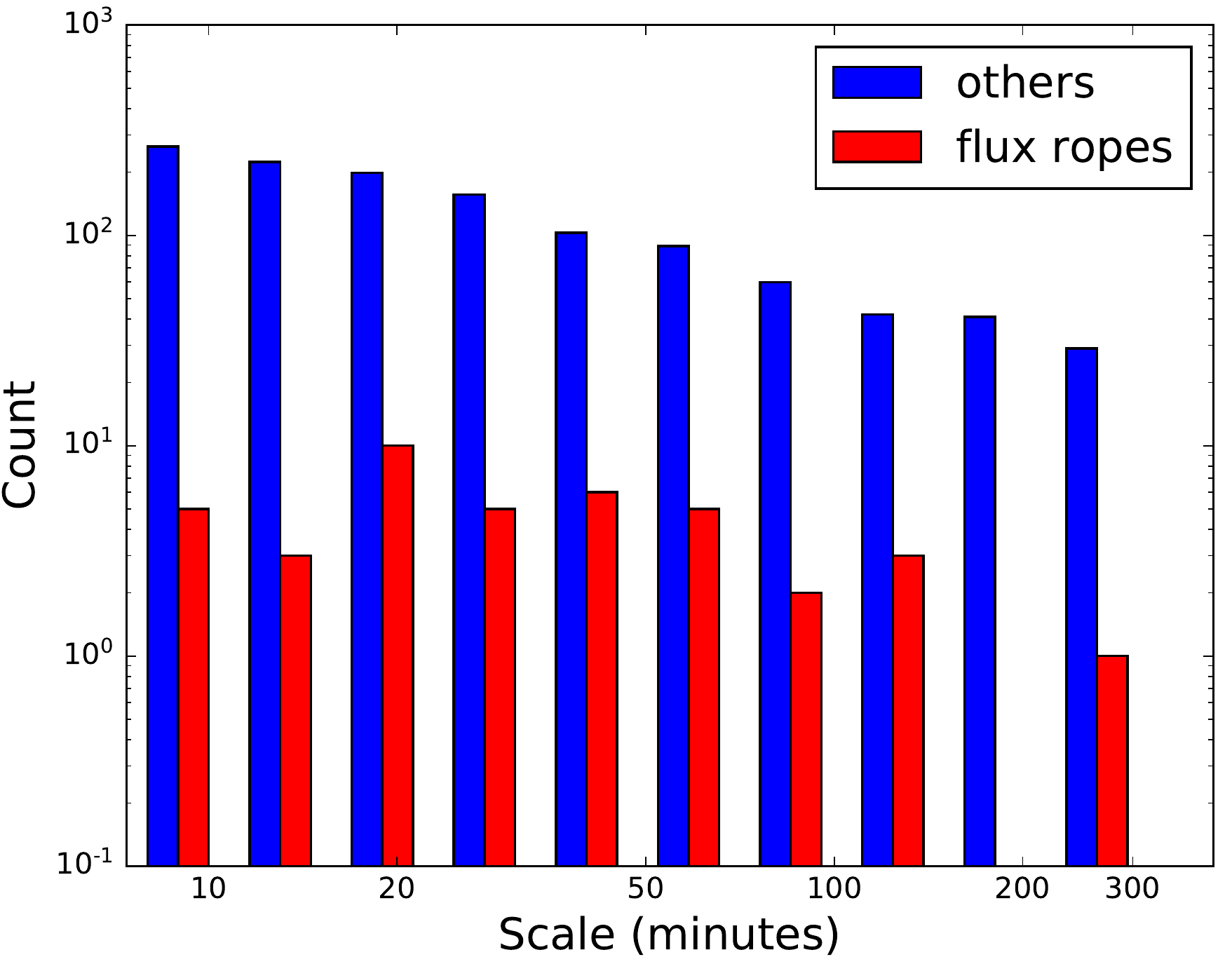}
\caption{Histogram of the scales of the identified structures with $|\sigma_m| \geq 0.7$.} \label{fig:histogram}
\end{figure}

A basic statistical analysis is performed on the scales of the identified structures, and the results are shown in Figure \ref{fig:histogram}. We create 10 bins uniformly in logarithmic scale between 7 and 300 minutes. Here, the structures that satisfy our flux rope criteria are plotted as red bars; these correspond to the ones that fall into the rectangular region in Figure \ref{fig:psp}. Other structures are plotted as blue bars; these include mostly Alfv\'enic structures. Figure \ref{fig:histogram} shows that most magnetic flux rope structures have duration smaller than 100 minutes, while there are more large-scale Alfv\'enic structures. Only the ICME event on 2018 November 12 has a scale larger than 200 minutes. Due to the limited number of structures detected, especially for flux ropes, we cannot draw a clear conclusion regarding the scale distribution for the period we considered.

\section{Conclusions}\label{sec:conclusion}
In this paper, the magnetic field and plasma data from PSP's first obit are analyzed. Using a wavelet analysis technique, we construct spectrograms of the magnetic helicity, cross helicity, and residual energy for the time period between 2018 October 22 and 2018 November 21. Two examples of the spectral characteristics are shown in the paper, including the ICME event observed on 2018 November 12 (264 minutes in duration) and the small-scale flux rope observed on 2018 November 13 (116 minutes in duration). We apply the analysis to the entire month-long dataset and 1245 structures are identified in total based on the criterion of large normalized magnetic helicity $|\sigma_m| > 0.7$. We then classify these events as magnetic flux ropes or Alfv\'enic structures according to their cross helicity and residual energy. The former are structures with small cross helicity and highly negative residual energy, while the latter have close-to-zero residual energy. By further limiting cross helicity ($|\sigma_c|<0.4$) and residual energy ($\sigma_r < -0.5$), we find 40 magnetic flux rope events with scales ranging between 8 minutes and $\sim$ 300 minutes. The parameters of these flux rope events are tabulated in the paper. A statistical analysis suggests that magnetic flux ropes are mostly found in the slow solar wind, while the fast solar wind is dominated by Alfv\'enic structures. These findings are in nice agreement with previous statistical studies. For example, \cite{Yu2014} found that many SFRs are more likely to be observed in the slow rather than fast solar wind. Unlike the large-scale ICME event, the proton temperature $T_p$ inside SFRs is not significantly less than the expected $T_p$. Thus, a low $T_p$ or plasma beta is not a robust signature of SFRs, although they are generally considered to be an essential features of large-scale ICMEs (e.g., the ICME event identified in Figure \ref{fig:CMEparameter}). Our findings is also consistent with the composite quasi-2D-slab turbulence model of the slow solar wind \citep{Zank2017}, and the slab turbulence of the fast solar wind. We also show the scale distribution of the detected structures, and find most of the detected flux ropes have relatively small scale $<$100 minutes. It should be noted that our criteria used for identifying magnetic flux ropes are based on an analysis of two example events and may not account for all SFRs that might be present in the data, but only include the most probable candidates. More detailed study on SFRs identified by some other technique (e.g., GS reconstruction) could help refine the thresholds and cross-check the results. 

In conclusion, our study presents, for the first time, the observational evidence of small-scale magnetic flux ropes within 0.3 au as measured by PSP during its first encounter. The connection between the coherent structures and particle acceleration or heating of plasma is yet to be understood and needs further investigation.

\acknowledgments
\section*{\leftline{Acknowledgement}}
We acknowledge the partial support of the NSF EPSCoR RII-Track-1 Cooperative Agreement OIA-1655280, 
NASA grants NNX08AJ33G, Subaward 37102-2, NNX14AC08G, NNX14AJ53G, A99132BT, RR185-447/4944336 and NNX12AB30G. 
Q.H. acknowledges 80NSSC18K0623, 80NSSC19K0276, and NSF AGS-1650854 for support. The SWEAP and FIELDS investigation and this publication are supported by the PSP mission under NASA contract NNN06AA01C. 
L.L.Z. and L.A. thanks K.E.Korreck, A.W.Case, and M.Stevens for their kind hospitality while visiting the Smithsonian Astrophysical Observatory (SAO).

\end{document}